\documentstyle[12pt]{article}
\title{ Quantum determinism \\ 
from quantum general covariance }
\author{Hrvoje Nikoli\'c \\
Theoretical Physics Division, Rudjer Bo\v{s}kovi\'{c} Institute, \\
P.O.B. 180, HR-10002 Zagreb, Croatia \\
{\normalsize e-mail: hrvoje@thphys.irb.hr} \\
\makebox[1in]{} \\
}
\date{\today}
\begin{document}
\maketitle
\begin{abstract}
The requirement of general covariance of quantum field theory (QFT) 
naturally leads to quantization based on the manifestly covariant 
De Donder-Weyl formalism. To recover the standard 
noncovariant formalism without violating covariance, fields 
need to depend on time in a specific deterministic manner. 
This deterministic evolution of quantum fields is recognized as
a covariant version of the Bohmian hidden-variable interpretation of QFT.
\end{abstract}
\vspace*{0.5cm}

\vspace*{0.9cm}

The reconcilation of quantum theory with general theory of relativity
is still an unsolved problem. It is very likely that the successful 
reconcilation requires a radical reformulation of the basic principles 
of relativity, or that of quantum theory, or both. One obvious difference 
between quantum theory and general relativity
is that quantum theory, in contrast with general relativity, is an 
{\em undeterministic}\footnote{Of course, the ideterminism 
is explicit only in the interpretation of quantum theory related to the 
problem of measurement. We note that the recent progress in understanding
the phenomenon of decoherence through the interaction with the 
environment shed much light on the problem of measurement in quantum theory,
but that this problem is still considered unsolved \cite{schl}.}  
theory.
Most attempts towards 
the reconcilation start from the assumption that quantum gravity, 
just as any quantum theory,
should also be an undeterministic theory. 
However, in contrast with this mainstream 
quantum-undeterministic paradigm,
't Hooft suggests that a fundamental theory
that reconciles quantum theory with general relativity 
should be a {\em deterministic hidden-variable} theory \cite{thooft}.
As a support for this idea, 
in this essay we argue that a deterministic hidden-variable
formulation of quantum field theory (QFT) 
naturally emerges from the requirement that quantum field theory 
should be general-covariant. In simple terms, 
restoring one classical property 
(general covariance) in quantum theory 
automatically restores another one (determinism).
Our discussion is based on recent results first presented 
by us in \cite{nikolepjc}, which, however, are logically
independent of the arguments presented by 't Hooft \cite{thooft}.

Canonical quantization of fields apparently contradicts 
theory of relativity because the formalism of canonical 
quantization requires a choice of a special time coordinate.
It is known that this fact does not destroy the covariance
of QFT with respect to Lorentz transformations \cite{bjor2}.
However, what about general coordinate transformations?
(In the rest of the paper, by the term ``covariant" we mean 
``general covariant".) 
QFT can be written in a covariant form by 
introducing states that are not functions of time, but 
functionals of an {\em arbitrary} hypersurface 
\cite{tom,schw,oec,rov,dopl,nikolpla}. (The hypersurface is 
often, but not always, restricted to be timelike.)
In this way, there is no preferred foliation of spacetime,
so quantization of fields is covariant. 
However, there is one problem with such a formalism. 
Without a preferred foliation of spacetime, the notion 
of a particle in QFT does not have an invariant meaning
\cite{ful,birdev,davies,rovpart}. Conversely, if a preferred 
foliation of spacetime is allowed, then the notion of 
a particle in QFT can be introduced in a local covariant 
manner \cite{nikolplb,nikolijmpd,nikolgrg}. But then the 
preferred foliation breaks the covariance of the quantization 
of fields themselves, so, again, the full covariance 
of the theory is lost.

Is it possible to have {\em both} quantum fields and particles 
described in a covariant manner? It {\em is} possible if 
a preferred foliation of spacetime is generated {\em dynamically}.
What we need is
a dynamical vector quantity $R^{\mu}$, the direction 
of which determines the preferred foliation. Since classical 
field theory is manifestly covariant without a 
dynamical preferred foliation, this 
vector should {\em not} be just another dynamical field that 
can be treated either as a classical or a quantum field. 
Instead, it should be a quantity that is inherently 
related to the quantization formalism itself. 
Thus, the natural starting point 
is to consider a scalar quantity of the conventional quantum 
formalism that can be promoted to a vector by recognizing 
that the original scalar is actually a time-component 
of a vector. 
The most obvious such quantity is the canonical 
momentum $\pi=\partial{\cal L}/\partial(\partial_0\phi)$ 
(where, for simplicity, $\phi(x)$ is a real scalar field).
Clearly, the canonical momentum
is a time-component of the momentum vector
\begin{equation}\label{mom}
\pi^{\mu}=\frac{\partial{\cal L}}{\partial(\partial_{\mu}\phi)} .
\end{equation}  
With the momentum (\ref{mom}), one naturally associates the covariant 
De Donder-Weyl Hamiltonian (see, e.g., \cite{kas,riet} and 
references therein)
\begin{equation}
{\cal H}(\pi^{\alpha},\phi)=\pi^{\mu}\partial_{\mu}\phi -{\cal L} .
\end{equation}   
One can also introduce the covariant De Donder-Weyl Hamilton-Jacobi equation
\cite{kas,riet}
\begin{equation}\label{HJc}
{\cal H}\left( \frac{\partial S^{\alpha}}{\partial\phi}, \phi\right)
+ \partial_{\mu}S^{\mu}=0,
\end{equation}
supplemented with the equation that governs the $x$-dependence 
of the field
\begin{equation}\label{eom}
\partial^{\mu}\phi=\frac{\partial S^{\mu}}{\partial\phi}.
\end{equation}
In a noncovariant language, equation (\ref{eom}) can be written 
as two independent equations 
\begin{equation}\label{eom2}
\partial^{0}\phi=\frac{\partial S^{0}}{\partial\phi}, \;\;\;
\partial^{i}\phi=\frac{\partial S^{i}}{\partial\phi}.
\end{equation}
The first equation in (\ref{eom2}) represents the ``dynamics" 
and corresponds to an analogous equation in the ordinary 
noncovariant Hamilton-Jacobi formalism. The second equation in (\ref{eom2})
says nothing about the time dependence of the 
field, so it is merely a ``kinematic" equation. However, it is clear 
that if one requires covariance, then the two equations in (\ref{eom2})
are {\em not} independent. Instead, it is crucial that 
{\em if the ``kinematic" part of (\ref{eom2})
is valid and if covariance is required, then the 
``dynamic" part of (\ref{eom2}) must also be valid}.
Another crucial point is the following: In order to recover 
the ordinary noncovariant Hamilton-Jacobi equation from the covariant 
Hamilton-Jacobi equation (\ref{HJc}), the quantity $S^i$ should be 
eliminated via the ``kinematic" part of (\ref{eom2}) \cite{kan,nikolepjc}.
Therefore, the ``kinematic" part of (\ref{eom2})
{\em must} be valid.

Now consider quantization. In the conventional noncovariant quantization
based on the Schr\"odinger picture, 
one replaces the ordinary noncovariant Hamilton-Jacobi equation 
with the corresponding noncovariant Schr\"odinger equation.
The Schr\"odinger state $\Psi=Re^{iS/\hbar}$ is desribed by two 
real functionals $R$ and $S$. Similarly, in the covariant approach 
based on the covariant Hamilton-Jacobi equation (\ref{HJc}), 
the quantum state is described by two real {\em vectors} 
$R^{\mu}$ and $S^{\mu}$ \cite{nikolepjc}. 
(See also \cite{kan,hip} for a different approach.)
In contrast with $S^{\mu}$, the vector $R^{\mu}$ does not possess 
a classical counterpart. Thus, it appears natural to identify $R^{\mu}$
as the vector that dynamically generates the preferred foliation of
spacetime \cite{nikolepjc}. With such a preferred foliation, 
the correspondence between covariant states and conventional states
takes the form
\begin{equation}\label{SR}
S=\int_{\Sigma} d\Sigma_{\mu}S^{\mu}, \;\;\;
R=\int_{\Sigma} d\Sigma_{\mu}R^{\mu},
\end{equation} 
where the integration is taken over a hypersurface $\Sigma$
that belongs to the dynamically preferred foliation.
For other details of the formalism, we refer the reader to \cite{nikolepjc}.

For the subject of this essay, the crucial point is the following.
The quantum analog of the covariant Hamilton-Jacobi equation (\ref{HJc}) 
must be compatible with the conventional Schr\"odinger equation.
The conventional Schr\"odinger equation can be recovered when 
$R^{\mu}=(R^0,0,0,0)$. However,
just as in the clasical case, the conventional Schr\"odinger equation
can be recovered only if
the ``kinematic" part of (\ref{eom2}) is valid. As we have 
seen, the requirement of covariance then implies that  
{\em the ``dynamic" part of (\ref{eom2}) must also be valid}.
This ``dynamic" part says that, in the Schr\"odinger picture, 
the field has a {\em deterministic dependence on time}.
On the other hand, in the conventional formulation of 
the Schr\"odinger picture of QFT, there is 
no equation that attributes a deterministic time dependence 
to the field. Instead, such a time dependence of the field 
corresponds to the {\em Bohmian interpretation} of QFT 
\cite{bohm2,bohmrep,holrep,holbook,nikfpl1,nikfpl2}.
In the literature, the Bohmian interpretation is viewed as a 
deterministic hidden variable theory postulated only
for interpretational purposes. Here, the Bohmian interpretation 
is {\em not postulated}, but {\em derived\footnote{Perhaps
it would be more honest to say that we found a new evidence
for the viability of the Bohmian interpretation, rather
than claim that we derived it in a strict sense.}  
from the requirement 
of covariance}. (Similarly, the Bohmian interpretation of strings 
can be derived from the world-sheet covariance \cite{nikolstring}.)
This, together with the results of
\cite{nikol1,nikol2} on relativistic first quantization,
suggests that it is Bohmian mechanics that might constitute the missing
bridge between quantum theory and relativity.    
The specific theory based on the De Donder-Weyl formalism proposed
here may also have measurable consequences based on the fact 
this theory actually generalizes standard QFT by allowing new 
types of quantum nonlocalities \cite{nikolepjc}. 
As more generic predictions based on a preferred foliation of
spacetime that defines a preferred notion of particles
we mention the measurable predictions on the 
Unruh effect \cite{nikmpla} and semiclassical gravity \cite{nikcc}.

At the end, we note that quantization based on the 
covariant De Donder-Weyl Hamiltonian leads to covariant quantization 
not only of matter fields in a fixed curved background
(in this case, some of the vectors above should be redefined 
as vector {\em densities} \cite{nikolepjc}), but also of 
gravity itself \cite{nikolepjc}. In the case of gravity, all ten components 
$g_{\mu\nu}$ of the metric tensor are quantized. 
In contrast with the conventionl noncovariant 
Wheeler-DeWitt approach to quantum gravity 
(see, e.g., \cite{padm,kuc2,ish} and references therein), 
there is no problem of time 
in the covariant approach. The consistency
with the classical noncovariant Hamiltonian constraint 
is obtained through the use of the covariant Bohmian equations of motion.  
This is how our covariant deterministic method of quantization 
resolves some deep conceptual problems of quantum gravity
by making quantum gravity more similar to classical gravity.

\section*{Acknowledgments}

This work was supported by the Ministry of Science and Technology of the
Republic of Croatia. 
%under Contract No.~0098002.

\end{document}